\documentclass{bioinfo}
\usepackage{url}

\copyrightyear{}
\pubyear{}

\usepackage{graphicx}
\usepackage{natbib}
\usepackage[normalem]{ulem}
 \usepackage{color}
 \definecolor{darkgreen}{rgb}{0,0.6,0}
 \definecolor{orange}{rgb}{0.99,0.257,0}
\newcommand{\rot}[1]{{\color{black} #1}}

\usepackage{amssymb}
\usepackage{amsfonts,amsmath}
\usepackage{epstopdf}
\usepackage{epsfig}
\usepackage{units}
\newlength{\FIGSIZE} 
\newcommand{\taua}{\tau_a}
\newcommand{\lrrund}[1]{\left( #1 \right)}

\newcommand{\vth}{v_\text{T}}

\newcommand{\pd}[2]{\frac{\partial#1}{\partial#2}}
\newcommand{\tr}[1]{#1^{\text{T}}}

\newcommand{\cv}{C_\text{V}}
\renewcommand{\vec}[1]{\mathbf{#1}}
\newcommand{\taum}{\tau_{\text{m}}}

\usepackage{fancyhdr,hyperref,lastpage}
\fancypagestyle{firststyle}
{
 \fancyhf{}
 
 \rhead{\sf Front. Comput. Neurosci. 2017; 7:164}
  \rfoot{\thepage/\pageref{LastPage}}
  
\lfoot{\sf Frontiers in Computational Neuroscience $\vert$ \href{http://dx.doi.org/10.3389/fncom.2013.00164}{doi:10.3389/fncom.2013.00164} $\quad$ October 26, 2013 }
}

\allowdisplaybreaks[1]
\begin{document}

\firstpage{1}

\title[Interval correlations in adapting neurons]{Patterns of interval correlations in neural oscillators with adaptation}
\author[Schwalger {et~al}]{Tilo Schwalger\,$^{1,2}$\footnote{to whom correspondence should be addressed, tilo@pks.mpg.de}\, and Benjamin Lindner\,$^{1,2}$}
\address{$^{1}$Bernstein Center for Computational Neuroscience, Haus 2, Philippstr 13, 10115 Berlin, Germany\\
$^{2}$Department of Physics, Humboldt Universit\"at zu Berlin, Newtonstr 15, 12489 Berlin, Germany}

\history{}

\editor{}
\maketitle
\thispagestyle{firststyle}



\begin{abstract}

\section{}
Neural firing is often subject to negative feedback by adaptation
currents. These currents can induce strong correlations among the time
intervals between spikes.  Here we study analytically the interval
correlations of a broad class of noisy neural oscillators with
spike-triggered adaptation of arbitrary strength and time scale.  Our
weak-noise theory provides a general relation between the correlations
and the phase-response curve (PRC) of the oscillator\rot{, proves
  anti-correlations between neighboring intervals for adapting neurons
  with type I PRC and identifies a single order parameter that
  determines the qualitative pattern of correlations. Monotonically
  decaying or oscillating correlation structures can be related to
  qualitatively different voltage traces after spiking, which can be
  explained by the phase plane geometry.} At high firing rates, the
long-term variability \rot{of the spike train} associated with the
cumulative interval correlations becomes small, independent of model
details. Our results are verified by comparison with stochastic
simulations of the exponential, leaky, and generalized
integrate-and-fire models with adaptation.

\section{Keywords:}
spike-frequency adaptation, non-renewal process, serial correlation
coefficient, phase-response curve, integrate-and-fire model,
long-term variability
\end{abstract}


\section{Introduction}
The nerve cells of the brain are complex physical systems. They
generate action potentials (spikes) by a nonlinear, adaptive, and
noisy mechanism.  In order to understand signal processing in single
neurons, it is vital to analyze the sequence of the interspike
intervals (ISIs) between adjacent action potentials.  There is
experimental evidence accumulating that the spiking in many cases is
{\em not} a renewal process, i.e. a spike train with mutually
independent ISIs, but that intervals are typically correlated over a
few lags \rot{\citep{LowTei92,RatNel00,NeiRus01,NawBou07,EngLSG08} (further reports are reviewed in \citep{FarStru09,AviCha11})}. These
correlations are a basic statistics of any spike train with important
implications for information transmission and signal detection in
neural systems \citep{RatNel00,ChaLon01,ChaLin04,AviCha11} and
man-made signal detectors \citep{NikSto12}. They are often
characterized by the serial correlation coefficient (SCC)
\begin{equation}
  \label{eq:scc-def}
  \rho_k=\frac{\langle (T_i-\langle T_i\rangle)(T_{i+k}-\langle T_{i+k}\rangle)\rangle}{\langle (T_i-\langle T_i\rangle)^2\rangle},
\end{equation}
where $T_{i}$ and  $T_{i+k}$ are two ISIs lagged by an integer $k$ 
and $\langle\cdot\rangle$ denotes  ensemble averaging. \rot{ISI correlations
can be induced via correlated input to the neural dynamics, e.g. in the form
of external colored noise \citep{MidCha03,Lin04}, intrinsic noise from
ion channels with slow kinetics \citep{FisSch12}, or stochastic
narrow-band input \citep{NeiRus01,NeiRus05,BauSch13}.}

\rot{Another} ubiquitous mechanism for ISI correlations are slow feedback
processes mediating spike-frequency adaptation
\citep{ChaLon00,BenLon05,LiuWan01} -- a phenomenon describing the
reduced neuronal response to slowly changing stimuli
\citep{BenHer03,GabKra06}. In the stationary state, these adaptation
mechanisms are typically associated with short-range correlations with
a negative SCC at lag $k=1$ and a reduced Fano factor as demonstrated
by several numerical \citep{GeiGol66,Wan98,LiuWan01,BenMal10} and
analytical studies \citep{Urd11a,FarMul11,SchFis10,SchLin10}. The
correlation structure of adapting neurons can show qualitatively
different patterns, ranging from monotonically decaying correlations
to damped oscillations when plotted as a function of the lag
\citep{RatNel00}. Because ISI correlations shape spectral measures
\citep{ChaLin04}, they bear implications for neural computation in
general.  However, a simple theory that predicts and explains possible
correlation patterns is still lacking.

In this article, we present a relation between the ISI correlation
coefficient $\rho_k$ and a basic characteristics of nonlinear neural
dynamics, the {\em phase-response curve} (PRC).  The PRC quantifies
the advance (or delay) of the next spike caused by a small
depolarizing current applied at a certain time after the last spike
\rot{\citep{Erm96}}.  For neurons which integrate up their input
(integrator neurons), the PRC is positive at all times (type I \rot{PRC}) whereas neurons, which show subthreshold resonances
(resonator neurons), possess a PRC that is partly negative (type II
\rot{PRC}) \rot{\citep{Erm96,Izh05,ErmTer10}}.
  Below we show that resonator
neurons possess a richer repertoire of correlation patterns than
integrator neurons do.

\section{Results}

\subsection{Model}

Spike frequency adaptation can be modeled by Hodgkin-Huxley type
neurons with a depolarization-activated adaptation current
\citep{Wan98,ErmPas01,BenHer03}.  However, the spiking of such
conductance-based models can in many instances be approximated by
simpler multi-dimensional integrate-fire (IF) models that are equipped
with a spike-triggered adaptation current \rot{\citep{Tre93,Izh03,BreGer05}}; adapting
IF models perform excellently in predicting spike times of real cells
under noisy stimulation \citep{GerNau09}. Here, we consider a
stochastic nonlinear multi-dimensional IF model for the membrane
potential $v$, $N$ auxiliary variables $w_j$ ($j=1,\dots, N$) and a
spike-triggered adaption current $a(t)$:
\begin{subequations}
  \label{eq:model}
  \begin{align}
    \label{eq:gen-oscill-adap}
    \dot v&=f_0(v,\vec{w})+\mu-a+\xi(t),\\
    \dot w_j&=f_j(v,\vec{w}),\\\label{eq:adap-current}
    \taua\dot a&=-a+\taua\Delta\sum_i\delta(t-t_i).
  \end{align}  
\end{subequations}
The membrane potential $v(t)$ is subject to weak Gaussian noise
$\xi(t)$ with $\langle\xi(t)\xi(t')\rangle=2D\delta(t-t')$ and noise
intensity $D$. The dynamics is complemented by a spike-and-reset
mechanism: whenever $v(t)$ reaches a threshold $\vth$, a spike is
registered at time $t_i=t$ and $v(t)$ and
$\vec{w}(t)=\tr{[w_1(t),\dotsc,w_{N}(t)]}$ are reset to
$v(\rot{t_i^+})=0$ and $\vec{w}(\rot{t_i^+})=\vec{w}_r$ \rot{(where
  $t_i^+$ denotes the right-sided limit $t\rightarrow t_i+0$). At the
  same time,} $a(t)$ suffers a jump by $\Delta\ge 0$ as seen from
Eq.~\eqref{eq:adap-current}, which resembles high-threshold adaptation
currents \citep{Wan98,LiuWan01}. The constant input current $\mu$ is
assumed to be sufficiently large to ensure ongoing spiking even in the
absence of noise. \rot{Note that the model is nondimensionalized by
  measuring time in units of the membrane time constant $\taum\sim
  10$~ms and voltage in units of the distance between reset and
  spike-initiating potential (a typical value is $15$~mV). In
  particular, the adaptation time constant $\taua$ is measured
  relative to $\taum$ and the unit of the firing rate is
  $\taum^{-1}\sim 100$~Hz. }

An important special case, the adaptive exponential integrate-and-fire
model \citep{BreGer05} \rot{with purely spike-triggered adaptation and
  a white noise current with constant mean} is illustrated in
Fig.1. \rot{It assumes an exponential nonlinearity $f_0(v)=-\gamma
  v+\gamma\Delta_T\exp[(v-1)/\Delta_T]$ \citep{FouHan03,BadLef08} and
  corresponds to $N=0$.}  Time courses of $v(t)$ and $a(t)$ are shown
in Fig.1a1,b1 for two distinct correlation patterns possible in this
model.  The ISIs $T_i=t_i-t_{i-1}$ are obtained as differences between
subsequent spiking times $t_i$. The sequence $T_i,T_{i+1},T_{i+2}$
displays patterns of {\em short-long-long} (Fig.1a1) and {\em
  short-long-short} (Fig.1b1), corresponding to a negative SCC, which
decays monotonically with the lag $k$ (Fig.1a3) or to an SCC
oscillating with $k$ (Fig.1b3). In the following, we develop a theory
to analyze these and other correlation patterns possible in
multi-dimensional adapting IF models.

\begin{figure}[t]
  \centering
  \includegraphics[width=8.6cm]{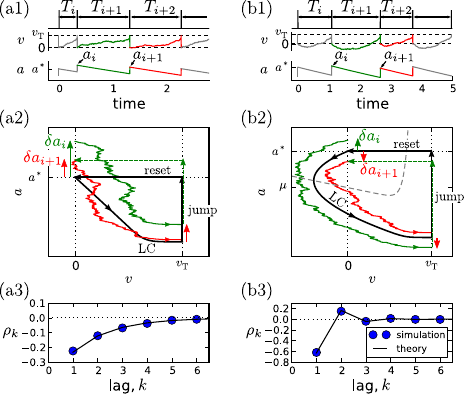}
  \caption{Correlation patterns in the adaptive exponential IF model
    with $\taua=10,\gamma=1, \Delta_T=0.1$, $\vth=2$,
    $D=0.1$. Adaptation is weak ($\Delta=1, \mu=15$) in (a) and strong
    ($\Delta=10, \mu=80$) in (b).  Membrane voltage $v(t)$ and
    adaptation variable $a(t)$ with ISI sequences $\{T_i\}$ and peak
    adaptation values $\{a_i\}$ are shown in (a1,b1)\rot{; time is in units
    of the membrane time constant $\taum$}.  Colored pieces of
    trajectories in the phase plane $(v,a)$ in (a2,b2) correspond to
    the respective colors in (a1,b1). \rot{The} deterministic limit
    cycle (LC)\rot{, determined by the initial (post-spike) values
      $v=0$, $a=a^*$,} is indicated by a thick black line. For weak
    adaptation (a2) a short ISI $T_{i}$ causes positive deviations
    $\delta a_{i}=a_i-a^*$ and $\delta a_{i+1}=a_{i+1}-a^*$ of peak
    values leading to long ISIs $T_{i+1}$ and $T_{i+2}$ and, hence, to
    a negative ISI correlation at all lags (a3). Because of the
    qualitatively different limit cycle for strong adaptation (b2),
    deviations $\delta a_{i}$ and $\delta a_{i+1}$ differ in sign,
    yielding an oscillatory correlation pattern (b3).  }
  \label{fig:trajec}
\end{figure}

\subsection{General theory}
\label{sec:theory}

 In our model Eq.~\eqref{eq:model}, $a(t)$ is the only
variable that keeps a memory of the previous spike times thereby
inducing correlations between ISIs. Over one ISI the time course of
adaptation is an exponential decay, relating two adjacent peak
values $a_{i}=a(t_{i}^+)$ and $a_{i+1}=a(t_{i+1}^+)$ by
\begin{equation}
  \label{eq:amap-general}
  a_{i+1}=a_{i}e^{-T_{i+1}/\taua}+\Delta
\end{equation}
(Fig.~\ref{fig:trajec}a1,b1). We assume that in the deterministic case ($D=0$) our model has a
finite period $T^*$ (i.e. the model operates in the tonically firing
regime) and, hence, for $D=0$ the map \eqref{eq:amap-general} has a
stable fixed point
\begin{equation}
  \label{eq:a-peak-limit-cycle}
  a^*=\Delta/\left[1-\exp(-T^*/\taua)\right].
\end{equation}
The asymptotic deterministic dynamics can be interpreted as a
limit-cycle like motion in the phase space from the reset point to the
threshold and back by the instantaneous reset
(cf. Fig.1a2,b2). 

Weak noise will cause small deviations in the period $\delta
T_{i}=T_i-T^*\approx T_i-\langle T_i\rangle$ that are mutually
correlated with coefficient $\rho_k=\langle\delta T_i\delta T_{i+k}\rangle/\langle\delta T_i^2\rangle$. The peak adaptation values,
however, also fluctuate, $\delta a_{i}=a_i-a^*$, and both deviations
are related by linearizing Eq.~\eqref{eq:amap-general}:
\begin{equation}
  \label{eq:lin-map}
  \delta T_{i+1}=\frac{\taua}{a^*}\left(\delta a_{i}-e^{T^*/\taua}\delta a_{i+1}\right).
\end{equation}
A second relation between the small deviations can be gained by
considering how a small perturbation in the voltage dynamics affects
the length of the period. This effect is captured by the infinitesimal
phase response curve (PRC), $Z(t)$, $t\in(0,T^*)$
\citep{Izh05,ErmTer10} (see Sec.~\ref{sec:material-methods} for the
precise definition). During the interval $T_{i+1}$, the voltage
dynamics in Eq.~\eqref{eq:gen-oscill-adap} can be written as $\dot
v=f_0(v,\vec{w})+\mu-(a^*+\delta
a_i)e^{-(t-t_{i})/\taua}+\xi(t)$. Compared to the deterministic limit
cycle, the dynamics is perturbed by the weak noise and the small
deviation in the adaptation $\delta a_ie^{-(t-t_{i})/\taua}$ yielding
in linear response
\begin{equation}
  \label{eq:strc}
  \delta T_{i+1}=\int_0^{T^*}\mathrm{d}t\,Z(t)\lrrund{\delta a_{i}e^{-\frac{t}{\taua}}-\xi(t_{i}+t)}.
\end{equation}
Combining Eqs.~\eqref{eq:lin-map}, \eqref{eq:strc} we obtain the stochastic map
\begin{equation}
  \label{eq:stochastic-map-ai}
  \delta a_{i+1}=\alpha\vartheta \delta a_i+\Xi_i,
\end{equation}
where $\Xi_i=-\frac{\alpha
  a^*}{\taua}\int_0^{T^*}\mathrm{d}t\,Z(t)\xi(t_i+t)$ are uncorrelated
Gaussian random numbers and
\begin{equation}
  \label{eq:vartheta}
  \alpha=e^{-T^*/\taua},\qquad  \vartheta=1-\frac{a^*}{\taua}\int_0^{T^*}\mathrm{d}t\,Z(t)e^{-\frac{t}{\taua}}.
\end{equation}
Note that local stability of the fixed point $a^*$ requires that
$|\alpha\vartheta|<1$. The covariance $c_k=\langle \delta a_{i}\delta
a_{i+k}\rangle$ of the auto-regressive process
Eq.~\eqref{eq:stochastic-map-ai} can be calculated
by elementary means and
using Eq.~\eqref{eq:lin-map} 
we obtain for $k\ge 1$:
\begin{equation}
  \label{eq:scc}
\rho_k=-A(1-\vartheta)(\alpha\vartheta)^{k-1},\qquad A=\frac{\alpha(1-\alpha^2\vartheta)}{1+\alpha^2-2\alpha^2\vartheta}.
\end{equation}
In order to compute $\alpha$ and $\vartheta$ via
Eq.~\eqref{eq:vartheta}, we have to calculate $T^*$ and $Z(t)$ ($a^*$
then follows from Eq.~\eqref{eq:a-peak-limit-cycle}), which can be
done for simple systems analytically.

Our main result, Eqs.~\eqref{eq:vartheta},\eqref{eq:scc}, allows to
draw a number of general conclusions. It shows that the SCC is always
a geometric sequence with respect to the lag $k$ that can generate
qualitatively different correlation patterns depending on the value of
$\vartheta$ and thus on PRC and adaptation current. Because
$|\alpha\vartheta|<1$ and $0<\alpha<1$, the prefactor $A$ in
Eq.~\eqref{eq:scc} is always positive. Consequently, $\rho_1$ is
negative for $\vartheta<1$ and positive for $\vartheta>1$. Looking at
Eq.~\eqref{eq:vartheta}, we find that a positive PRC inevitably yields
$\vartheta<1$. This implies that adapting neurons with type I PRC
possess negative correlations between adjacent ISIs. Intuitively, a
short ISI causes in the following on average a higher inhibitory
adaptation during the subsequent ISI. Such an inhibitory current
always enlarges the ISI in type I neurons -- hence, a short ISI is
followed by a long ISI.

The sign of higher lags is determined by the base of the power: for
$\vartheta>0$ correlations decay monotonically, whereas for
$\vartheta<0$ the SCC oscillates. Two special cases are $\vartheta=0$
with a negative correlation at lag 1 and vanishing correlations at all
higher lags and $\vartheta=1$ where all correlations
vanish. \rot{Overall, we find five basic patterns corresponding to
  $-\alpha^{-1}<\vartheta<0$, $\vartheta=0$, $0<\vartheta<1$,
  $\vartheta=1$ and $1<\vartheta<\alpha^{-1}$. These basic patterns
  cover all interval correlations discussed in previous theoretical
  studies \citep{SchLin10,Urd11a}. Our geometric formula generalizes
  the theory for the perfect IF model with adaptation \citep{SchFis10}
  to more realistic, nonlinear multi-dimensional IF models with
  adaptation.}

The cumulative effect of the correlations can be described by the sum
over all $\rho_k$, which determines the long-time limit of the Fano
factor and the low-frequency limit of the spike train power spectrum
(\rot{for a definition of these quantities,} see Sec.~\ref{sec:relat-betw-second}). Evaluating the geometric series
yields
\begin{equation}
  \label{eq:sumsrho}
  \sum_{k=1}^\infty\rho_k=-\frac{A(1-\vartheta)}{1-\alpha\vartheta}.
\end{equation}
This shows that adaptation in neurons with type I resetting
($\vartheta<1$) leads to a negative summed correlation and hence a
reduced long-term variability. Furthermore, at high firing rates
achieved by a strong input current $\mu$, the sum in
Eq.~\eqref{eq:sumsrho} can be approximated by
\begin{equation}
  \label{eq:limit-sumsrho}
  \sum_{k=1}^\infty\rho_k\simeq-\frac{1}{2}+\frac{1/2}{\left(1+\Delta \taua/\vth\right)^2},\qquad T^*\ll \taua.
\end{equation}
In particular, for strong adaptation ($\Delta\taua\gg\vth$) the sum is
only slightly larger than $-1/2$. Note that by virtue of the
fundamental relation
$\lim_{t\rightarrow\infty}F(t)=C_\text{V}^2\left(1+2\sum_{k=1}^\infty\rho_k\right)$
\citep{CoxLew66b} (see Sec.~\ref{sec:relat-betw-second}), the smallest
possible value for the sum is $-1/2$ in order to ensure the
non-negativity of the Fano factor $F(t)$.  At this minimal value the
long-term variability as expressed by the Fano factor vanishes even
for a non-vanishing ISI variability as quantified by the coefficient
of variation $C_\text{V}$. \rot{The latter quantity can also be
  estimated using the weak-noise theory: From
  Eq.~\eqref{eq:stochastic-map-ai} one can calculate the variance of
  $a_i$ and using Eq.~\eqref{eq:lin-map} an approximation for
  $\cv^2\approx\langle\delta T_i^2\rangle/{T^*}^2$ can be obtained:
\begin{equation}
  \label{eq:cv}
  \cv^2=2D\frac{1+\alpha^2-2\alpha^2\vartheta}{[1-(\alpha\vartheta)^2]{T^*}^2}\int_0^{T^*}\mathrm{d}t\,[Z(t)]^2.
\end{equation}}

\subsection{One-dimensional IF models with adaptation} 
In the simplest case ($N=0$, $f_0(v,\vec{w})=f(v)$) the PRC reads
\begin{equation}
  \label{eq:one-dim-prc}
Z(t)=Z(T^*)\exp\Bigl[\int_t^{T^*}\mathrm{d}t'\,f'(v_0(t'))\Bigr],
\end{equation}
where $v_0(t)$ is the limit cycle solution and
$Z(T^*)=[f(\vth)+\mu-a^*+\Delta]^{-1}$ is the inverse of the velocity
$\dot v_0(T^*)$ at the threshold, which is always positive. Thus, the
PRC is positive for all $t\in(0,T^*)$, i.e. one-dimensional IF models
show type I behavior. From our general considerations, this implies a
negative SCC at lag $k=1$. The sign of the correlations at higher lags
can be inferred from the sign of $\vartheta$, for which one can show
(Sec.~\ref{sec:material-methods}) that
\begin{equation}
  \label{eq:theta-1d}
  \vartheta=(f(0)+\mu-a^*)Z(0).
\end{equation}
Because $Z(0)>0$, the sign of $\vartheta$ is determined by the sign of
$f(0)+\mu-a^*$.  For \rot{weak adaptation such that $a^*<f(0)+\mu$
(achieved by a sufficiently small value of $\Delta$ or $\taua$,
Fig.~\ref{fig:trajec}a), we will have} $\vartheta>0$ and a negative
correlation at all lags (Fig.~\ref{fig:trajec}a3). In this case, a
short ISI occurring by fluctuation will cause a positive deviation
$\delta a_i$ (Fig.~\ref{fig:trajec}a2, green arrow). Geometrically, it
is plausible that such a positive deviation causes a likewise positive
deviation $\delta a_{i+1}$ in the subsequent cycle
(Fig.~\ref{fig:trajec}a2, red arrow). Because a positive deviation is
associated with a long ISI, the initial short ISI is on average
followed by longer ISIs.

In marked contrast, \rot{for strong adaptation such that
  $a^*>f(0)+\mu$ (achieved by a sufficiently large value of $\Delta$
  or $\taua$)}, $\vartheta$ becomes negative and hence the SCC's sign
alternates with the lag. This alternation of the sign can be
understood by means of the phase plane. Let us again consider a
positive deviation $\delta a_i$ due to a short preceding ISI
(Fig.~\ref{fig:trajec}b2, green arrow). Because $\dot
v_0(0)=f(0)+\mu-a^*<0$, the neuron is reset above the $v$-nullcline
and hence hyperpolarizes at the beginning of the interval, i.e. the
trajectory makes a detour into the region of negative voltage
\rot{(corresponding to a ``broad reset'' in \citet{NauMar08})}. A
positive deviation $\delta a_i$ leads to a larger detour (green
trajectory) causing a sign inversion and hence a negative deviation
$\delta a_{i+1}$ (Fig.~\ref{fig:trajec}(b2), red arrow). Because a
positive (negative) deviation corresponds on average to a long (short)
ISI, the alternation of $\delta a_i$ also entails an alternation of
the ISI correlations.  \rot{Thus, the distinction between monotonic and
alternating patterns relates to a qualitative distinction of the
voltage trace after resetting (cf. ``sharp'' vs. ``broad'' resets in
\citet{NauMar08}).}

\begin{figure}[t]
  \centering
  \includegraphics[width=8.6cm]{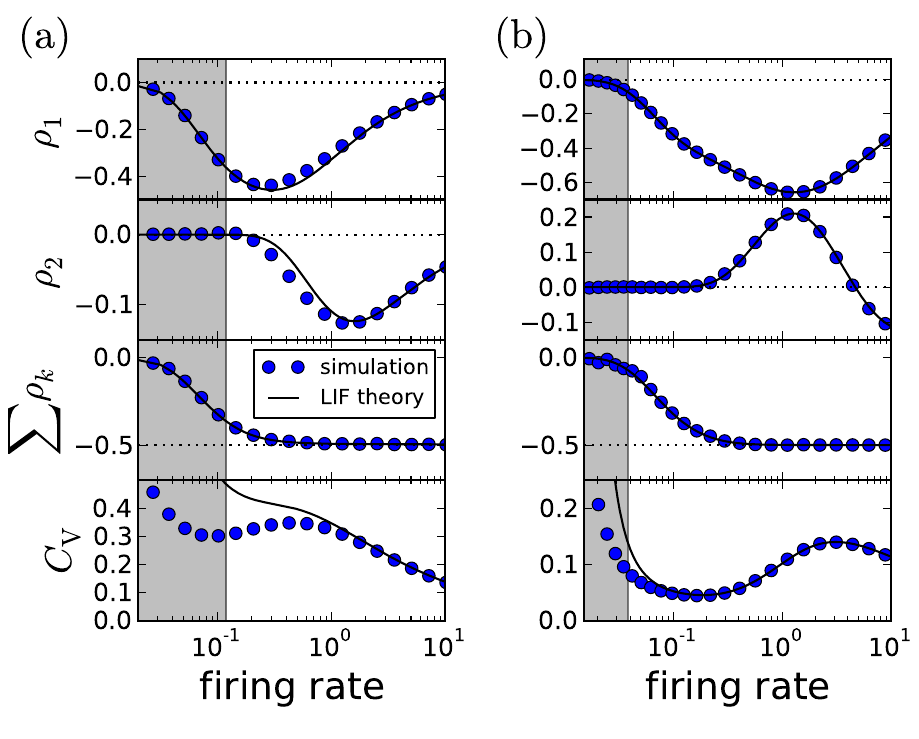}
  \caption{ISI correlations \rot{and coefficient of variation (CV)} of
    the adapting LIF model vs. firing rate $1/\langle
    T_i\rangle\approx 1/T^*$, where the rate is varied by increasing
    $\mu$. \rot{The gray-shaded area corresponds to the
      fluctuation-driven regime ($\mu<\gamma\vth$), where the theory
      is expected to fail.} The panels display (from top to bottom)
    $\rho_1$, $\rho_2$, the sum $\sum_{k=1}^{m}\rho_k$ \rot{and the
      CV} for simulation (circles, $m=100$) and theory (solid lines,
    $m\to\infty$).  (a) Moderate adaptation: $\Delta=1$, (b) strong
    adaptation: $\Delta=10$.  Both: $\gamma=1$, $\taua=10$, $D=0.1$,
    $\vth=1$. \rot{Note that the firing rate is given in units of the
      inverse membrane time constant $\taum^{-1}$.}}
  \label{fig:fig2}
\end{figure}

As demonstrated in Fig.\ref{fig:trajec}a3,b3, our theory works well
for the adapting exponential integrate-and-fire model. We next
demonstrate the validity of our approach over a broad range of firing
rates (Fig.~\ref{fig:fig2}) for another important 1D model, the
adapting leaky integrate-and-fire model \citep{Tre93} for which
$f(v)=-\gamma v$ and
\begin{equation}
Z(t)=\exp[\gamma(t-T^*)]/(\mu-\gamma \vth-a^*+\Delta)
\end{equation}
(here $T^*$ has still to be determined from a transcendental
equation).  Changing the firing rate by varying the input current
$\mu$, we find a good agreement for the first two correlation
coefficients and the sum of all $\rho_k$\rot{; the approximation of
  the CV shows deviations from simulation results when the input
  current $\mu$ becomes weak small (approaching the fluctuation-driven
  regime).} In accordance with previous findings
\citep{Wan98,LiuWan01,NesMal10,BenMal10,SchFis10,SchLin10,Urd11a}, the
first correlation coefficient $\rho_1$ displays a minimum
corresponding to strong anti-correlations between adjacent
intervals. The correlations at lag $2$ can be positive for a finite
range of firing rates if the adaptation strength is sufficiently large
(Fig.~\ref{fig:fig2}(b)), whereas for moderate adaptation we find a
negative $\rho_2$ at all firing rates (Fig.~\ref{fig:fig2}(a)). In
both cases, however, the sum of SCCs approaches a value close to
$-1/2$ for high firing rates as predicted by
Eq.~\eqref{eq:limit-sumsrho} (Fig.~\ref{fig:fig2}, bottom). This is
strikingly similar to experimental data from weakly electric fish, in
which some electro-receptors display a monotonically decaying SCC and
some show an oscillatory SCC \citep{RatNel00} but all cells exhibit a
sum close to $-1/2$ \citep{RatGoe04}. \rot{Finally, we notice a local
maximum of the CV for some suprathreshold current $\mu$ -- an effect
that has been described by \citet{NesNeg08}.}

\subsection{Generalized integrate-and-fire model with adaptation} %
Different correlation patterns become possible if we consider a type
II PRC, which is by definition partly negative and can lead to a
negative value of the integral in Eq.~\eqref{eq:vartheta}, and hence
to $\vartheta\ge 1$. This corresponds to a non-negative SCC at lag
$1$, which is infeasible in the one-dimensional case. To test the
prediction $\rho_1\ge 0$, we study the generalized integrate-and-fire
(GIF) model \citep{BruHak03} with spike-triggered adaptation. This
model is defined by $f_0(v,w)=-\gamma v-\beta w$ and
$f_1(v,w)=(v-w)/\tau_w$. Using the method described in
Sec.~\ref{sec:material-methods}, the PRC is obtained as
\begin{equation}
  \label{eq:Zgif}
  Z(t)=\frac{e^{\frac{\nu}{2}(t-T^*)}\left[\cos(\Omega(t-T^*))-\frac{1-\tau_w\gamma}{2\tau_w\Omega}\sin(\Omega(t-T^*))\right]}{\mu-\gamma\vth-\beta
w_0(T^*)-a^*+\Delta}
\end{equation}
where $\nu=\gamma+1/\tau_w$,
$\Omega=\sqrt{\frac{\beta+\gamma}{\tau_w}-\frac{\nu^2}{4}}$ and
$w_0(t)$ is one component of the deterministic limit-cycle solution
$[v_0(t),w_0(t),a_0(t)]$ that we calculated numerically.

\begin{figure}[t]
  \centering
  \includegraphics[width=8.6cm]{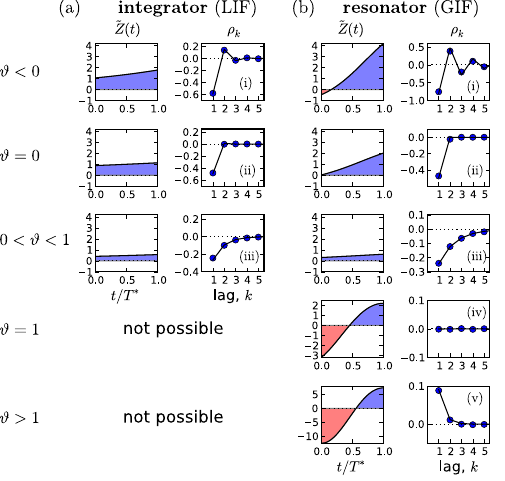}
  \caption{Possible correlation patterns and corresponding PRCs
    (solid
      lines: theory, symbols: simulations of Eq.~\eqref{eq:model}).  For the
    adapting LIF model (a), $\vartheta<1$ and only three qualitative
    different cases are possible.  The adapting GIF model (b) exhibits
    the full repertoire of correlation patterns because the PRC can be
    partly negative and $\vartheta$ can attain values from its entire
    physically meaningful interval $[-1/\alpha,1/\alpha]$. The value
    of $\vartheta$ and hence the type of correlation pattern is set by
    the integral over the weighted PRC $\tilde
    Z(t)=Z(t)e^{-\frac{t}{\taua}}\frac{a^*T^*}{\taua}$, shown in left
    panels.  LIF parameters: $D=0.1$, $\taua=2$, (i) $\mu=20$,
    $\Delta=10$, (ii) $\mu=20$, $\Delta=4.47$, (iii) $\mu=5$,
    $\Delta=1$.  GIF parameters: (i) $\mu=10$, $\beta=3$,
    $\taua=10$. (ii) $\mu=11.75$, $\beta=3$, $\taua=10$. (iii)
    $\mu=20$, $\beta=1.5$, $\taua=10$. (iv) $\mu=2.12$, $\beta=1.5$,
    $\taua=1$, $\Delta=10$. (v) $\mu=1.5$, $\beta=1.5$, $\taua=1$,
    $\Delta=9$ $D=10^{-5}$.  Unless stated otherwise, $\gamma=1$,
    $\Delta=1$, $\tau_w=1.5$, $D=10^{-4}$, $w_r=0$.}
  \label{fig:gif}
\end{figure}

In Fig.~\ref{fig:gif}b we demonstrate that all possible correlation
patterns can be realized in the GIF model and that the predicted SCCs
agree quantitatively well in theory and model simulations (for
comparison, see the SCC for the LIF in Fig.~\ref{fig:gif}a). To each
distinct pattern belongs a range of $\vartheta$ (Fig.~\ref{fig:gif},
left), determined by the area under the weighted PRC $\tilde
Z(t)=\frac{a^*}{\taua}e^{-\frac{t}{\taua}}Z(t)$. The function $\tilde
Z(t)$ (left column in Fig.~\ref{fig:gif}a,b) illustrates, why an
adapting GIF neuron can show vanishing (Fig.~\ref{fig:gif}b(iv)) or
even {\em purely positive} ISI correlations (Fig.~\ref{fig:gif}b(v)).
In case of type II resetting, inhibitory input can {\em shorten} the
ISI because of the negative part in the PRC; here inhibition acts like
an excitatory input.  Consequently, a short ISI will induce a stronger
inhibition (adaptation) that now causes a likewise short interval and
results thus in a positive correlation between adjacent ISIs. Also,
the shortening effect of the adaption current in the early negative
phase of the PRC can be exactly balanced by the delaying effect of the
late positive phase of the PRC (pseudo-renewal case, in which the area
under $\tilde{Z}$ is zero).

\section{Discussion}
We have found a general relation between two
experimentally accessible characteristics: the serial interval
correlations and the phase response curve of a noisy neuron with
spike-triggered adaptation. The theory predicts distinct correlation patterns like
short-range negative and oscillatory correlations that have been
observed in experiments \citep{RatNel00,NawBou07} and in simulation
studies of adapting neurons \citep{ChaLon00,LiuWan01}. 
Beyond negative and oscillatory correlations, we have found, however,
that resonator neurons with spike-frequency adaptation can exhibit
purely positive ISI correlations or a pseudo-renewal process with
uncorrelated intervals. Adaptation currents that are commonly
associated with negative ISI correlations
\rot{\citep{Wan98,LiuWan01,ChaLon01,ChaPak03,NesMal10,BenMal10} can thus
induce a rich repertoire of correlation patterns.}  Despite the
multitude of patterns, there is a universal limit for the cumulative
correlations at high firing rates (cf. Eq.~\eqref{eq:limit-sumsrho}),
which shows that the long-term variability \rot{of the spike train} is
in this limit always reduced in agreement with experimental studies
\citep{RatGoe04}.

Our analytical results apply to arbitrary adaptation strength and time
scale but \rot{require that} (i) the noise is weak \rot{and white},
(ii) the deterministic dynamics shows periodic firing with equal ISIs
(i.e. a limit-cycle exists) and (iii) the adaptation current is purely
spike-triggered with (iv) a single \rot{exponential decay
  time}. Regarding the weak-noise assumption, we found from numerical
simulations quantitative agreement with our theory for values of the
coefficient of variation (CV) up to 0.4, which is, for instance,
typical for neurons in the sensory periphery
\citep{RatNel00,NeiRus04,VogHen05}. This holds even in the
subthreshold regime at low CVs\rot{, where the deterministic system
  does not follow a limit cycle. In this case, $T^*$ has to be}
replaced by the mean ISI. Moreover, we found qualitative agreement
even for moderately strong noise with values of the CV up to 0.8,
which is typical for cortical non-bursting neurons in vivo (e.g. Fig.3
in \citep{SofKoc93}).

\rot{In the absence of a deterministic limit-cycle, i.e. in} the
fluctuation-driven regime at high CVs, different mathematical
approaches have to be employed, such as those based on a
hazard-function formalism
\citep{MulBue07,NesMal10,SchLin10,FarMul11}. \rot{Furthermore, for
  some parameter sets, we also observed repeat periods of the
  deterministic system that involved multiple ISIs corresponding to a
  periodic ISI sequence with $T_i=T_{i+n}$, where the smallest period
  is $n\ge 2$. Such cases can realize bursting \citep{NauMar08}, which
  we did not consider in the present study. However, we expect that
  these parameter regimes yield interesting correlation patterns
  because already in the noiseless case a periodic ISI sequence
  exhibits correlations between ISIs.}

Regarding the last two assumptions, it seems that the analytical
derivation cannot be easily extended to the cases of adaptation
currents activated by the subthreshold membrane potential
(``subthreshold adaptation''
\citep{ErmPas01,BreGer05,PreSej08,DeeKro12}) and multiple-time-scale
adaptation \citep{PozNau13}.
\citet{ErmPas01} have shown that the inclusion of subthreshold
adaptation can lead to type II PRCs, which according to our theory
could \rot{qualitatively} change the correlation patterns. \rot{An
  adaptation dynamics depending on the subthreshold membrane potential
  also involves a fluctuating component because $v$ is
  noisy. According to \citet{SchFis10}, this stochasticity could
  contribute positive correlations. The combined effect of
  spike-triggered, subthreshold and stochastic adaptation currents on
  the sign of the SCC is not clear.}  

\rot{The important cases of the fluctuation-driven regime and
  multiple-time-scale adaptation have been recently analyzed with
  respect to the first-order spiking statistics including the
  stationary firing rate as well as the mean response to
  time-dependent stimuli \citep{Ric09,NauGer12}. However, the
  second-order statistics, which describes the fluctuations of the
  spike train (``neural variability'',
  cf. Sec.~\ref{sec:relat-betw-second}) and which limits the
  information transmission capabilities of neurons, is still poorly
  understood theoretically in these cases. How adaptation shapes
  second-order statistics in the cases of multiple adaptation time
  scales and fluctuation-driven spiking is an interesting topic for
  future investigations.}

As an outlook we sketch, how our theory could be used to constrain
unknown physiological parameters by measured SCCs and PRCs. For
instance, from the mean ISI we can estimate $T^*=\langle
T\rangle$. Furthermore, knowing
$\rho_1=-A(\alpha,\vartheta)(1-\vartheta)$ as well as the ratio
$\rho_2/\rho_1=\alpha\vartheta$ one can eliminate $\vartheta$ and
solve for $\alpha$. This allows to estimate the unknown adaptation
time constant $\taua=-T^*/\ln\alpha$ and the amplitude of the
adaptation current
\begin{equation}
  \label{eq:alpha-extract}
  a^*=\left.\frac{\taua}{\alpha}\left(\alpha-\frac{\rho_2}{\rho_1}\right)\middle/\int_0^{T^*}\mathrm{d}t\,Z(t)e^{-\frac{t}{\taua}}\right..
\end{equation}
Although experimental PRCs are notoriously noisy \citep{Izh05}, the
integral over $Z(t)$ determining our estimate of $a^*$ is less
error-prone. Combining our approach with advanced estimation methods
for the PRC \citep{GalErm05}, may thus provide an alternative access to hidden physiological parameters using only spike time statistics. 

\begin{methods}
\section{Material and Methods}
\label{sec:material-methods}
\subsection{Phase-response curves of adapting IF models}
We use the phase-response curve $Z(t')$ to characterize the shift of
the {\em next} spike following a small current pulse applied at a
given ``phase'' $t'\in[0,T^*]$ of an
ISI. 
More
precisely, let us assume that the last spike occurred at time
$t_0=0$. Then, the next spike time $t_1$ of the perturbed limit cycle
dynamics $\dot v=f_0(v,\vec{w})+\mu-a+\epsilon\delta(t-t')$, $v(0)=0$,
$\vec{w}(0)=\vec{w}_r$, $a(0)=a^*$, $0<t'\le T^*$ will be shifted by
some amount $\delta T(t',\epsilon)=t_1-T^*$. The infinitesimal PRC can
be defined as the limit
\begin{equation}
  \label{eq:prc-def}
  Z(t')=-\lim_{\epsilon\rightarrow 0}\frac{\delta T(t',\epsilon)}{\epsilon},
\end{equation}
where the sign has been chosen such that a spike advance ($\delta
T<0$) due to a positive stimulation ($\epsilon>0$) leads to a positive
PRC. The definition of $Z(t)$ by the shift of the next spike differs
from the PRC that describes the asymptotic spike shift but is
equivalent to the so-called ``first-order PRC'', which is often
measured in experiments \citep{NetSch12}.

\subsubsection{Adjoint equation and boundary conditions}

The PRC can be computed using the adjoint method (see
e.g. \citet{ErmTer10}). To this end, the dynamics is linearized about
the $T^*$-periodic limit cycle solution
$\vec{y}_0(t)=[v_0(t),\vec{w}_0(t),a_0(t)]$. The linearized limit-cycle
dynamics $\vec{y}(t)=\vec{y}_0(t)+\delta\vec{y}(t)$ corresponding to Eq.~\eqref{eq:model} is given by
\begin{equation}
  \label{eq:lin-dyn}
  \dot{\delta\vec{y}}=A(t)\delta\vec{y}
\end{equation}
with the Jacobian matrix
\begin{equation}
  \label{eq:linearized-model}
A(t)=
\begin{pmatrix}
  \pd{f_0}{v}&\pd{f_0}{w_1}&\dots&\pd{f_0}{w_N}&-1\\
  \tau_1^{-1}\pd{f_1}{v}&\tau_1^{-1}\pd{f_1}{w_1}&\dots&\tau_1^{-1}\pd{f_1}{w_N}&0\\
  \hdotsfor[2]{5}\\
  \tau_N^{-1}\pd{f_N}{v}&\tau_N^{-1}\pd{f_N}{w_1}&\dots&\tau_N^{-1}\pd{f_N}{w_N}&0\\
0&\hdotsfor[2]{2}&0&-\taua^{-1}
\end{pmatrix}
\end{equation}
evaluated at $v=v_0(t),\vec{w}=\vec{w}_0(t)$. The linear response of
the ISI to perturbations of the limit-cycle dynamics in an arbitrary
direction is given by the vector
$\vec{Z}(t)=[Z(t),Z_{w_1}(t),\dotsc,Z_{w_N}(t),Z_a(t)]^\text{T}$,
where the first component is equal to the PRC defined above. This
vector satisfies the adjoint equation
$\dot{\vec{Z}}=-A^\text{T}\vec{Z}$ ($A^\text{T}$ denotes the transpose
of $A$) with the normalization condition $\dot
v_0(t)Z(t)+\dot{\vec{w}}_0(t)\vec{Z}_{w}(t)+\dot a_0(t)Z_a(t)=1$. The
remaining $N+1$ boundary conditions are obtained by the following
consideration: On the limit cycle $\Gamma$, a phase
$\phi:\Gamma\rightarrow[0,T^*]$ can be introduced in the usual way by
inverting the map $t\mapsto \vec{y}_0(t)$ and setting
$\phi=t$. Because we are interested in the shift of the {\em next}
spike, it is useful to define the isochrons (sets of equal phase) as
the sets of all points in phase space that will lead to the same first
spike time. Put differently, phase points belonging to the same
isochron will have their first threshold crossing in synchrony. As a
consequence, the threshold hyperplane defined by the condition
$v=\vth$ is a special isochron corresponding to the phase
$\phi=T^*$. Note that this definition of the phase implies that the
reset line defined by the condition $v=0,\vec{w}=\vec{w}_r$ does
generally {\em not} correspond to $\phi=0$ but to positive phases if
$a<a^*$ and negative phases if $a>a^*$. Thus, off-limit-cycle
trajectories suffer a phase jump upon reset. Close to the threshold,
the isochrons are parallel to the threshold, and thus, a perturbation
perpendicular to the $v$-direction does not change the phase. This
insensitivity implies the boundary conditions
$Z_{w_1}(T^*)=\dotsc=Z_{w_N}(T^*)=Z_a(T^*)=0$. Note that a definition
of the PRC based on the asymptotic spike shift would require periodic
boundary conditions \citep{LadAug12}.

From the above considerations, it becomes clear that the PRC $Z(t)$
can be computed for $t\in[0,T^*]$ by solving the system
\begin{equation}
  \label{eq:prc-comput}
  \begin{pmatrix}
    \dot Z\\\dot Z_{w_1}\\\vdots\\\dot Z_{w_N}
  \end{pmatrix}
  =-
\begin{pmatrix}
  \pd{f_0}{v}&\tau_1^{-1}\pd{f_1}{v}&\dots&  \tau_N^{-1}\pd{f_N}{v}\\
  \pd{f_0}{w_1}&\tau_1^{-1}\pd{f_1}{w_1}&\dots&\tau_N^{-1}\pd{f_N}{w_1}\\
  \vdots&\vdots&\ddots&\vdots\\
  \pd{f_0}{w_N}&\tau_1^{-1}\pd{f_1}{w_1}&\dots&\tau_N^{-1}\pd{f_N}{w_N}
\end{pmatrix}
  \begin{pmatrix}
    Z\\Z_{w_1}\\\vdots\\Z_{w_N}
  \end{pmatrix}
\end{equation}
subject to the boundary conditions
\begin{align}
  \label{eq:bcs}
  Z(T^*)&=\frac{1}{\dot v_0(T^*)}=\frac{1}{f_0(\vth,\vec{w}_0(T^*))+\mu-a^*+\Delta},\\
  Z_{w_k}(T^*)&=0,\qquad k=1,\dotsc,N.
\end{align}
The PRC with respect to $a$ is determined by
\begin{equation}
  \label{eq:za}
  \dot Z_a=\frac{1}{\taua}Z_a+Z(t),\qquad Z_a(T^*)=0.
\end{equation}
The matrix in Eq.~\eqref{eq:prc-comput} is again evaluated on the
limit cycle at $v=v_0(t),\vec{w}=\vec{w}_0(t)$ and is therefore
time-dependent. An analytic solution of Eq.~\eqref{eq:prc-comput} is
possible for one-dimensional models with adaptation ($N=0$) or general
linear IF models although in most cases the deterministic period $T^*$
still has to be computed numerically.

\subsubsection{One-dimensional case}

In the case $N=0$, the PRC satisfies the equation $\dot Z=-f'(v_0)Z$
with boundary condition \eqref{eq:bcs}. The solution is given by
Eq.~\eqref{eq:one-dim-prc}. In order to prove Eq.~\eqref{eq:theta-1d},
we compute $Z_a(t)$ from Eq.~\eqref{eq:za} yielding
\begin{equation*}
Z_a(t)=e^{\frac{t}{\taua}}\left(Z_a(0)+\int_0^tZ(t')e^{-\frac{t'}{\taua}}\,\mathrm{d}t'\right).  
\end{equation*}
Evaluation of this expression for $t=T^*$ leads to
$\vartheta=1+\frac{a^*}{\taua}Z_a(0)$. Finally, using the
normalization condition $(f(0)+\mu-a^*)Z(0)-\frac{a^*}{\taua}Z_a(0)=1$
yields Eq.~\eqref{eq:theta-1d}.

\subsection{Relation between second-order statistics of spike count,
  spike train and interspike intervals}
\label{sec:relat-betw-second}

A stationary sequence of spike times
$\{\dotsc,t_{i-1},t_i,t_{i+1},\dotsc\}$ is often characterized by the
statistics of the spike train $x(t)=\sum_{i}\delta(t-t_i)$, the spike
count $N(t)=\int_0^t\mathrm{d}t'\,x(t')$ or the sequence of ISIs
$\{T_i=t_i-t_{i-1}\}$. In particular, neural variability can be quantified
by the second-order statistics of these different descriptions as, for
instance, the spike train power spectrum
\begin{equation}
  \label{eq:psd}
  S(f)=\int\mathrm{d}\tau\,e^{2\pi i f \tau}\langle x(t)x(t+\tau)\rangle,
\end{equation}
the Fano factor
\begin{equation}
  \label{eq:fano}
  F(t)=\frac{\langle N(t)^2\rangle-\langle N(t)\rangle^2}{\langle N(t)\rangle},
\end{equation}
and the coefficient of variation $C_\text{V}=\sqrt{\langle
  (T_i-\langle T_i\rangle)^2\rangle}/\langle T_i\rangle$ and SCC
$\rho_k$ as defined in Eq.~\eqref{eq:scc-def}. These statistics are
connected by the fundamental relationship \citep{CoxLew66b} (see also
\citep{Vre10})
\begin{equation}
  \label{eq:fano-psd}
  \lim_{t\rightarrow\infty}F(t)=\langle T_i\rangle\lim_{f\rightarrow 0} S(f)=C_\text{V}^2\left(1+2\sum_{k=1}^\infty\rho_k\right).
\end{equation}
It shows that the summed SCC has a strong impact on the long-term
variability of the spike train. In particular, a negative sum yields a
more regular spike train on long time scales than a renewal process
with the same $C_\text{V}$.

\end{methods}

\section*{Acknowledgement}
This work was supported by Bundesministerium f\"ur Bildung und Forschung grant 01GQ1001A.



\end{document}